\documentclass[
    ,final            
  ]
  {aipproc}

\layoutstyle{6x9}

\def\etal{{\it et al.}}
\def\eg{{\it e.g.,}}

\begin{document}

\title{Quasar Black Hole Masses from Velocity Dispersions}

\classification{98.54.Aj; 98.62.Js}
\keywords      {black hole physics --- galaxies: active --- quasars: general}

\author{Gabriela Canalizo}{
  address={Dept. of Physics and Astronomy and IGPP, 
University of California, Riverside, CA 92521, USA}
}

\author{Margrethe Wold}{
  address={Institute of Theoretical Astrophysics, University of Oslo, N-0315 Oslo, Norway}
}

\author{Mariana Lazarova}{
  address={Dept. of Physics and Astronomy and IGPP, 
University of California, Riverside, CA 92521, USA}
} 

\author{Mark Lacy}{
  address={Spitzer Science Center, 
California Institute of Technology, Pasadena, CA 91125, USA}
}

\begin{abstract}

Much progress has been made in measuring black hole (BH) masses in 
(non-active) galactic nuclei using the tight correlation between stellar 
velocity dispersions $\sigma$ in galaxies and the mass of their central BH.  
The use of this correlation in quasars, however, is hampered by the difficulty
in measuring sigma in host galaxies that tend to be overpowered by their very 
bright nuclei.  We discuss results from a project that focuses on 
$z\sim0.3$ quasars suffering from heavy extinction at shorter wavelengths.  
This makes it possible to obtain clean spectra of the hosts in the spectral 
regions of interest, while broad lines (like H$\alpha$) are still visible at 
longer wavelengths. We compare BH masses obtained from velocity dispersions to
those obtained from the BLR and thus probe the evolution of this relation and
BH growth with redshift and luminosity.  Our preliminary results show an 
offset between the position of our objects and the local relation, in the 
sense that red quasars have, on average, lower velocity dispersions than 
local galaxies.   We discuss possible biases and systematic errors that may 
affect our results.

\end{abstract}

\maketitle


\section{Background}

Black hole (BH) mass is believed to be one of the fundamental parameters 
that characterize quasar activity and much effort has been devoted to 
obtaining accurate BH masses for quasars and other AGN \citep[\eg][]{ho99}. 
In recent years, much progress has been made in measuring BH masses
in galactic nuclei, particularly with the remarkable discovery by 
\citet{geb00a} and \citet{fer00} of a tight correlation between stellar 
velocity dispersion in galaxies and the mass of their central BH 
(M$_{\rm BH}\propto\sigma^{n}$).
The use of this correlation to derive BH masses in AGN, 
however, is hampered by the difficulty in measuring velocity 
dispersions in host galaxies that tend to be overpowered by their 
very bright nuclei.  Nevertheless, the correlation has been shown
to be present at low redshift ($z<0.1$) in low luminosity AGN
(\eg\ BL Lac objects: \cite{barth03}; or Seyfert galaxies: 
\cite{geb00b}).  Seyfert galaxies at higher redshift ($z\sim0.36$ 
and $z\sim0.57$), however, appear to show an offset from the local relation
\citep[][and references therein]{Woo2008}.

It is not yet known whether the M$_{\rm BH}-\sigma$ correlation
holds for the highest luminosity AGN. A loose correlation
has been found by using the width of [O\,III] lines in active nuclei
\citep{nel00,shi03}, but the width of these lines is
dependent upon other parameters (outflows, radio luminosity, etc.) 
and therefore lead to a correlation with a large scatter.
BH masses derived from [O\,III] emission line widths can only
be accurate to within a factor of five at best \citep{bor03}.  More 
accurate determinations are necessary if we hope to use them to
disentangle some of the other fundamental relationships among
quasar parameters.

\section{Red quasars}

We are therefore carrying out a program to measure stellar velocity 
dispersions in quasar host galaxies.  We have 
selected a sample of $z<0.4$ $red$ quasars from 2MASS.  Red quasars 
are likely the dust obscured equivalent 
of the blue quasar population, and they have the advantage that
the nucleus is highly extincted at optical wavelengths, so that 
the contrast between the stellar flux from the host galaxy and 
that of the nucleus is increased.  Thus, the spectra of these 
objects show, at shorter 
wavelengths, stellar features that are useful to measure velocity 
dispersions and, at longer wavelengths, broad emission lines from which to 
obtain virial estimates of BH masses.

Our sample of 11 objects is drawn from \citet{mar03}.   We obtained deep, 
medium resolution spectroscopic observations with the Echelle Spectrograph 
and Imager (ESI) on the Keck II telescope.  We placed the slit through the 
center of the host galaxies in order to measure velocity dispersions 
of the bulges of the hosts. Figure~1 shows the spectrum of one of the 
objects in the sample demonstrating that they suffer little contamination 
from the nucleus at wavelengths shorter than H$\alpha$.

\begin{figure}
  \includegraphics[height=.4\textheight]{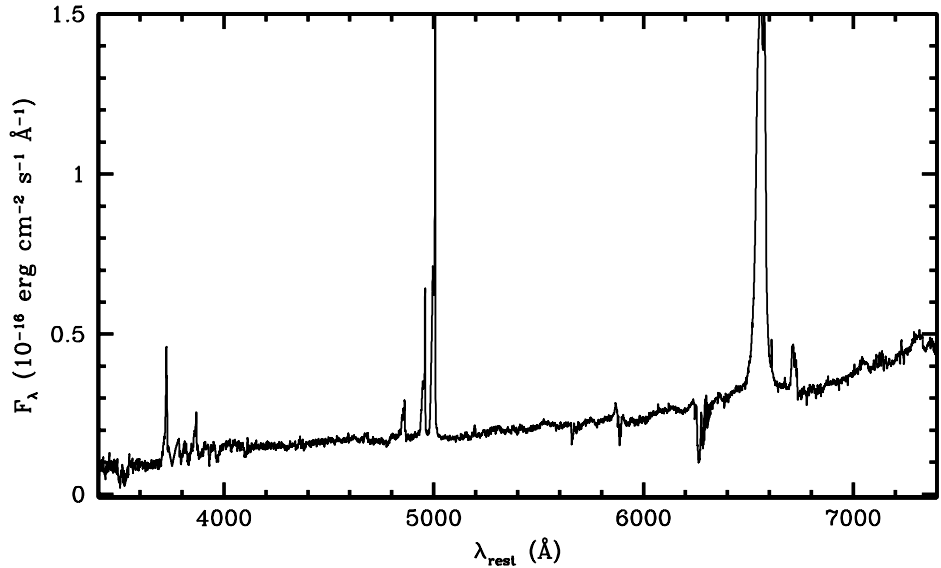}
\caption{\footnotesize  Keck ESI spectrum of a red 2MASS quasar.   
Although these objects show broad H$\alpha$ emission characteristic of quasars,
the spectra of the host galaxies suffer little contamination from the quasar 
at shorter wavelengths. 
}
\end{figure}

\subsection{Velocity Dispersions}

We first estimated the size of the stellar bulges by inspecting archival
$HST$/WFPC2 images (proposal ID 9057; PI D. Hines) and extracted spectra
from these regions.   
We measured velocity dispersions ($\sigma_{c}$) by fitting the spectra in the 
rest frame region between 5220 and 5550 \AA\ for each of the targets.  We 
used templates formed from the combination of spectra of stars of 
different spectral types observed also with ESI.  To these templates we 
added a small fraction of a continuum to simulate any potential 
contamination from the active nucleus.   We were able to obtain a reliable
$\sigma_{c}$ for eight of the targets, with typical errors at the 95\% 
confidence level of $\sim$ $\pm$ 20 km s$^{-1}$. 

\subsection{Black Hole Masses}

We estimated virial masses for the BHs in the sample by first fitting
the FWHM of the broad component of H$\alpha$, and then using the scaling 
relation given by \citet{kas00}.  In this relation, the size of the broad line
region is a function of the continuum luminosity at rest frame 5100 \AA.
However, as mentioned before, the quasar continuum suffers from heavy 
extinction in this spectral region.   To obtain the unobscured flux at 
5100 \AA, we used the following procedure: (1) We measured the flux of the
quasar at F814W in the HST images by fitting an empirical PSF to the nucleus.
In this way, we also determined the relative flux contribution from the host
galaxy and the quasar in the region covered by the slit.
(2) We scaled the Keck ESI spectrum to match the flux obtained from the HST
images.  (3) We fitted a reddened version of the SDSS composite quasar 
spectrum plus a reddened stellar population, using the relative contributions 
determined in (1), and varying the amount of reddening, guided by the measured
ratios of H$\alpha$/H$\beta$.  The E(B-V) that we measured in the sample 
varied from 0.5 to 2.   Measuring the extinction accurately is currently our 
main source of uncertainty and, until we perform more detailed fitting of
the spectra, our results are only tentative.


\begin{figure}[ht]
  \includegraphics[height=.51\textheight]{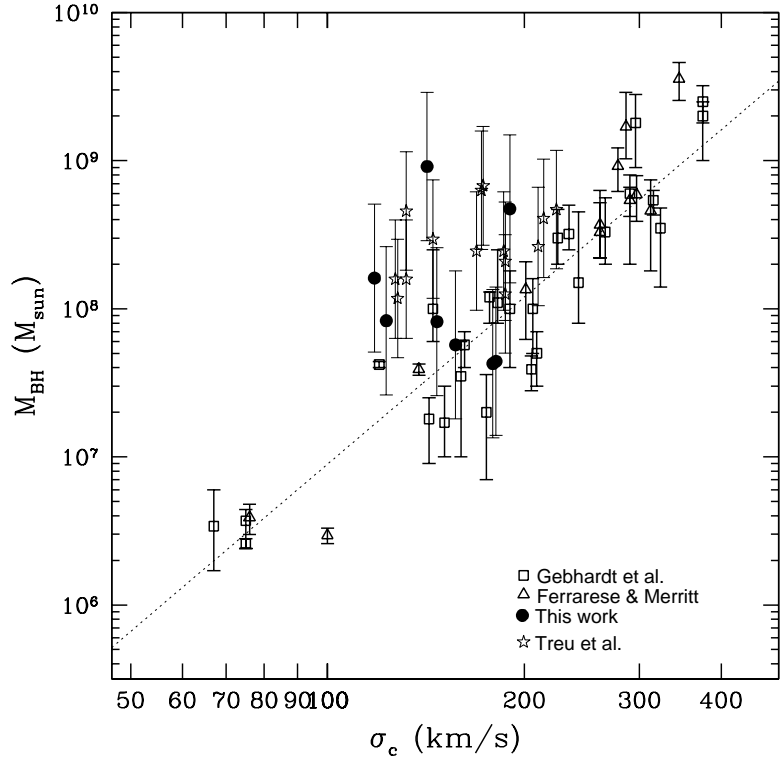}
  \caption{M$_{BH}$--$\sigma_{c}$ relation for red quasars. The objects in 
our sample are plotted as solid circles, along with local objects 
(open squares and triangles) and Seyfert galaxies at $z=0.37$ (open stars).
The dotted line marks the empirical relation derived for local objects.
}
\end{figure}

\subsection{The M$_{BH}$--$\sigma_{c}$ relation.} 

Preliminary results for the eight objects we measured are plotted as solid 
circles in Fig.~2, along with results from
local objects taken from \citet{geb00a} and \citet{fer00}, and Seyfert galaxies
at $z=0.37$ taken from \citet{treu04}.  While half of the objects have
positions consistent with the local relation, the other half seem to have an
offset in the sense that $\sigma$ has lower values for a given BH mass.  The
positions of these objects in the plot are more consistent with those of 
Seyfert galaxies at higher $z$ published by \citet{treu04}, who have found 
evidence for evolution in the relation from $z=0.57$ to the present 
\citep{Woo2008}.  The two objects that fall beneath the relation in Fig.~2 are 
indeed the ones with the lowest $z$ in our sample.  However, until we perform
more careful modeling to determine the nuclear extinction,
we can only speculate about this potential evidence pointing to evolution.


\begin{theacknowledgments}
Support for this project was provided by the NSF, 
under grant number AST 0507450.

\end{theacknowledgments}



\bibliographystyle{aipproc}   





\end{document}